\documentclass[journal]{ltxguide}[1995/11/28]
  \usepackage[logonly]{trace}
\usepackage{graphicx}
\usepackage{amsthm}
\usepackage{subfigure}

\title{Collaboration in sensor network research: an in-depth longitudinal analysis of assortative mixing patterns}

\author{Alberto Pepe\footnote{Alberto Pepe. Center for Embedded Networked Sensing. University of California, Los Angeles. Email: \texttt{apepe@ucla.edu}} \and Marko A. Rodriguez\footnote{Marko A. Rodriguez. T-5, Center for Nonlinear Studies. Los Alamos National Laboratory. Email: \texttt{marko@lanl.gov}}}

\begin{document}
\maketitle

\begin{abstract}
Many investigations of scientific collaboration are based on statistical analyses of large networks constructed from bibliographic repositories. These investigations often rely on a wealth of bibliographic data, but very little or no other information about the individuals in the network, and thus, fail to illustrate the broader social and academic landscape in which collaboration takes place. In this article, we perform an in-depth longitudinal analysis of a relatively small network of scientific collaboration ($N = 291$) constructed from the bibliographic record of a research center involved in the development and application of sensor network and wireless technologies. We perform a preliminary analysis of selected structural properties of the network, computing its range, configuration and topology. We then support our preliminary statistical analysis with an in-depth temporal investigation of the assortative mixing of selected node characteristics, unveiling the researchers' propensity to collaborate preferentially with others with a similar academic profile. Our qualitative analysis of mixing patterns offers clues as to the nature of the scientific community being modeled in relation to its organizational, disciplinary, institutional, and international arrangements of collaboration.
\end{abstract}

\newpage

\section{Introduction}
Scientific communities have large, well-established, and relatively well structured digital footprints which have increasingly been the focus of specialized research. These footprints, composed of scholarly publications and related artifacts, have been employed for bibliometric analyses involving coauthorship, citation, co-citation, acknowledgments, and other such indicators of scientific productivity and knowledge production \cite{furner}. Furthermore, as the majority of scientific publications are now available and consulted online, a number of recent analyses have progressively studied not only the production of scholarly artifacts but also their usage. Recent studies include: an analysis of the structure of readership networks in the field of education research \cite{Carolan200869} and a large-scale analysis of user request logs (clickstreams) gathered from scholarly web portals \cite{bollenings}. In general, these kinds of research are made possible because the scientific community is a large social institution that is open to investigation.

Coauthorship patterns are perhaps the most studied scholarly and scientific phenomena. Recent studies of coauthorship have analyzed the literature production within specific domains such as high energy physics \cite{mele}, genetic programming \cite{tomassini}, neuroscience \cite{braun}, nanoscience \cite{schummer:2004}, library science \cite{liu}, economics \cite{hollis:2001}, organizational science \cite{acedo:2006}, and psychology and philosophy \cite{cronin}. Similar analyses have also been comparative in nature and have explored social and normative differences of coauthorship behavior across different domains \cite{newman_pnas}. Moreover, an increasing number of studies of this kind have accounted for the evolving component of scientific collaboration \cite{barabasi2002,catanzaro:2004,Wagner:2005,snijders:2007}. 

More specifically, coauthorship patterns have been widely and actively studied from a social network analysis perspective for over two decades \cite{lievrouw:1987,Liberman:1998,Fenner:2007}. Most social network research involved with coauthorship is based upon this underlying concept: two individuals (nodes) are regarded as coauthors if they appear together in the author list of a publication (edge). This relational structure works reasonably well when investigating coauthorship patterns in scholarly arrangements where publications are authored by relatively small groups. It is true that some scientific domains have experienced an increase in the number of authors per publication making it impossible to discern the nature and extent of individual contributions to a publication \cite{cronin:2005}. A striking example of this phenomenon can be found in the domain of high-energy physics where author lists for a single publication often include tens or even hundreds of authoring researchers \cite{traweek}. For this reason, a number of recent studies of physics collaboration supplement traditional analytic techniques with more qualitative methods of survey research, i.e., directly asking authors to indicate the real nature of their contributions to a publication \cite{tarnow:2002,birnholtz:2006}. However, besides the singular case of high-energy physics, the vast majority of scholarly coauthorship networks incorporate collaboration circles of only a handful of authors per publication, suggesting that coauthorship activity can be adequately employed to construct a valid social network of collaboration \cite{newman:2000}.

In this article, we perform a temporal analysis of a coauthorship network constructed from the bibliographic record of a research center involved in the development and application of sensor network and wireless technologies. The network studied here is relatively small in size compared to networks generally analyzed in related research. The small size of the collaboration network results in a fundamental advantage: besides analyzing certain large-scale structural properties of the network, we can explore the social and academic arrangements in which collaboration patterns evolve, based on a set of node characteristics. 

The study of node characteristics can provide insights into the level of ``homophily'' in a social network, i.e. the tendency of individuals to create ties with similar others \cite[for a review]{homophily:cook2001}. The homophily principle describes how homogeneous a network is in terms of specific sociodemographic, behavioral, or interpersonal characteristics. For example, a high level of homophily in a friendship network indicates that individuals with certain characteristics---such as race, ethnicity, political beliefs, and educational background---tend to make friends with individuals with similar characteristics. Many studies of homophily are grounded in sociology and investigate patterns of homophily as well as their driving forces and their implications. An established method to measure mathematically the level of homophily in a network is by computing its ``assortativity'' (also ``assortative mixing''), i.e., the extent of mixing between similar nodes in a network \cite{newman:mixpatt2003}. While many different components of similarity can be investigated, the vast majority of large-scale studies of networks look at the mixing of node degree, i.e. how nodes with similar degree preferentially attach to one another. Mixing patterns, however, can also be calculated based on discrete node-specific characteristics. In studies of scholarly and scientific collaboration, examples of characteristics that have been investigated include research interests \cite{collab:havemann2006}, academic domain \cite{moody}, geographical location \cite{lorigo}, age group \cite{bonaccorsi:2003}, and country of origin \cite{rodrpepe} of the individuals in the network. Studying networks in terms of these node properties can offer insights into the mechanisms by which disciplinary, institutional, and spatial arrangements shape, and are shaped by, collaboration patterns.

The present study, exploratory in nature, ties a quantitative analysis of a network's assortativity structure to a qualitative explanation of the social and academic landscape in which such network is embedded. Scholars working with scientific collaboration networks are increasingly becoming interested in grounding their quantitative analyses in sociological theory. An example is the work of Kretschmer \& Kretschmer \cite{kretschmer} that investigates whether the distribution of co-author pairs' frequencies in a collaboration network can be regarded as a ``social Gestalt''. They derive a mathematical function to describe social Gestalts in networks and apply it on four large scale coauthorship networks (cumulative $N_{ij} > 2M$ co-author pairs) to explore the relationship between coauthorship and node-based author productivity. 

This study focuses on a much smaller collaboration network ($N = 291$). The small size of the network allows us to both perform a quantitative analysis of selected structural properties of the network and provide a a sociological explanation of our findings, based on an in-depth qualitative analysis of assortative mixing patterns. We demonstrate how certain social and academic dynamics, for example the emergence of new international collaborations or the inception of new inter-departmental efforts, have varying levels of control in the resulting topology and configuration of the scientific collaboration network.

\section{Present study}
This article presents the findings of a study of scientific collaboration at the Center for Embedded Networked Sensing (CENS).\footnote{The website of the Center for Embedded Networked Sensing (CENS) is available online at \texttt{http://research.cens.ucla.edu/}} CENS is a National Science Foundation (NSF) Science \& Technology Center funded in 2002. CENS supports interdisciplinary collaborations among faculty, students, and staff of five partner universities in Southern California: University of California, Los Angeles (UCLA); University of Southern California (USC); University of California, Riverside; California Institute of Technology (Caltech); and University of California, Merced. From 2005, CENS features a headquarter office base located within Boelter Hall at UCLA.

The mission of CENS research is to use sensor network technology to reveal previously unobservable phenomena. From its inception, CENS has developed and deployed sensor network devices for the study of a wide range of natural phenomena, such as seismic activity, fluid contaminant transport, and bird breeding behavior. Besides these pursuits in the natural sciences, the social and built environments have progressively become the focus of applied CENS research: sensing mobile systems are being employed for the study of public health, environmental protection, urban planning, and cultural expression.

The type of research conducted at CENS now spans a wide spectrum of disciplines and applications (from biology to seismology, from wireless telecommunications to statistics, from education to environmental science) requiring continuous cooperation among individuals that, otherwise, would probably not interact beyond the walls of traditional university departments and faculties. In such a diverse scholarly and scientific landscape, distributed collaboration on multi-disciplinary subjects constitute a fundamental leverage for scientific research. 

\subsection{Data collection}
Computing the range, population and configuration of an interdisciplinary, multi-institutional research center like CENS can be an arduous task. 
\begin{quote}
``In order to be recognisable as such, a system must be bounded in some way. However, as soon as one tries to be specific about the boundaries of a system, a number of difficulties become apparent'' \cite[p. 139]{cilliers-boundary}. 
\end{quote}
In the case presented in this article, these difficulties have to do with the inherently open and dynamic nature of modern science research centers. Unlike other types of organizational arrangements for which the boundaries are more or less evident (e.g. academic institutions and departments, corporate and government centers), many modern research centers and laboratories act as umbrella organizations with very flexible and blurry boundaries. CENS, for example, includes researchers from multiple institutions and disciplines. CENS scholars seamlessly interact with each other within and beyond institutional and departmental boundaries: collaboration patterns are ubiquitous and non-uniformly distributed. Researchers affiliated with CENS may also be affiliated with other research laboratories and perform interdisciplinary work on other projects and under different affiliations. Moreover, many CENS collaborations include researchers that are not officially affiliated with the center. In other words, the nature and extent of contribution to CENS collaboration depends on a number of organizational and scholarly factors, and is not solely restricted to individuals officially affiliated with the center. Under these conditions, what is the best way to construct a network that accurately captures scientific collaboration of this research center?

Previous environment-specific studies of collaboration have delineated the population under study by relying on publication data contained in an institutional repository \cite{acedo:2006} or domain-specific bibliographic databases \cite{liu} to mine patterns of coauthorship that take place within a given institution or academic domain, respectively. For the purpose of this article, we used a similar mechanism, thus including in our population not only CENS members, but also individuals that though not officially affiliated, have contributed to the production of CENS or CENS-related scholarly publications. 

We constructed a database of publications by assembling the items included yearly in the NSF Annual Reports, which contain the official list of documents published by CENS for a given fiscal year.\footnote{CENS Annual Reports are available online at \\ \texttt{http://research.cens.ucla.edu/about/annual\_reports/}} For the available reporting period (2002 through 2008) and for every publication, we collected the following information: a) author names, b) publication title, c) publication type, d) publication venue, and e) publication date. For the purpose of this article, we utilized author names and publication dates (years) to construct a coauthorship network, i.e. a network consisting of individuals (nodes) that are connected to each other (via edges) if they are recorded as authors on the same scholarly publication. The resulting bibliographic dataset consisted of 547 manuscripts (370 conference proceedings, 159 journal articles, 17 book chapters and 1 book), published over a period of 7 years (2001--2007). 

This bibliographic database was used to generate a weighted undirected network in which nodes represent authors and edges represent coauthorship activity among them. For example, if the present paper had to be included in this network, its authors (Pepe and Rodriguez) would become two distinct nodes, connected by an edge. In order to determine the weights between nodes, i.e. the strength of collaboration among coauthors, we used a weighting mechanism proposed by Newman \cite{newman_phys_II} by which the weight of the edge between nodes $i$ and $j$ is:
\begin{equation}
w_{ij} = \displaystyle\sum_{k} \frac{\delta^k_i \delta^k_j}{n_k-1},
\end{equation}
where $\delta^k_i$ is $1$ if author $i$ collaborated on paper $k$ (and zero otherwise) and $n_k$ is the number of coauthors of paper $k$. For the example above, the edge between authors Pepe and Rodriguez would have $w_{ij} = 1$. An article written by three authors (e.g., Pepe, Rodriguez, and Bollen) would result in three edges (Pepe-Rodriguez, Pepe-Bollen, and Rodriguez-Bollen), each one with $w_{ij} = 0.5$. And so on. As such, this weighting mechanism confers more weight to small and frequent collaborations, based on the assumptions that: i) publications authored by a small number of individuals involve stronger interpersonal collaboration than multi-authored publications, and ii) authors that have authored multiple papers together know each other better on average and thus collaborate more strongly than occasional coauthors \cite{newman_phys_II}. 

\begin{figure*}[htp]	
	\begin{center}
\subfigure[2002]{
	\includegraphics[width=0.40\textwidth]{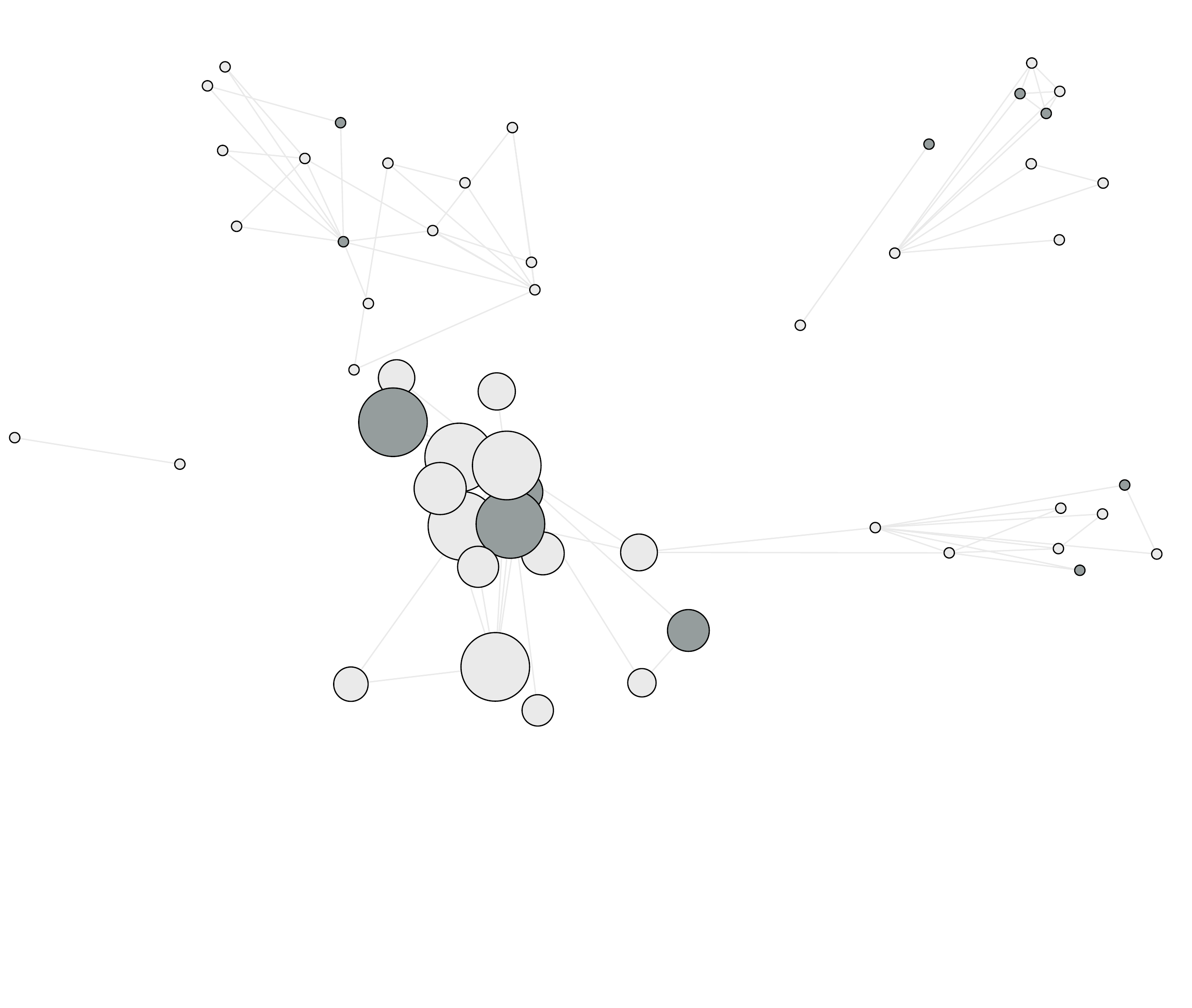}}
\subfigure[2003]{
	\includegraphics[width=0.40\textwidth]{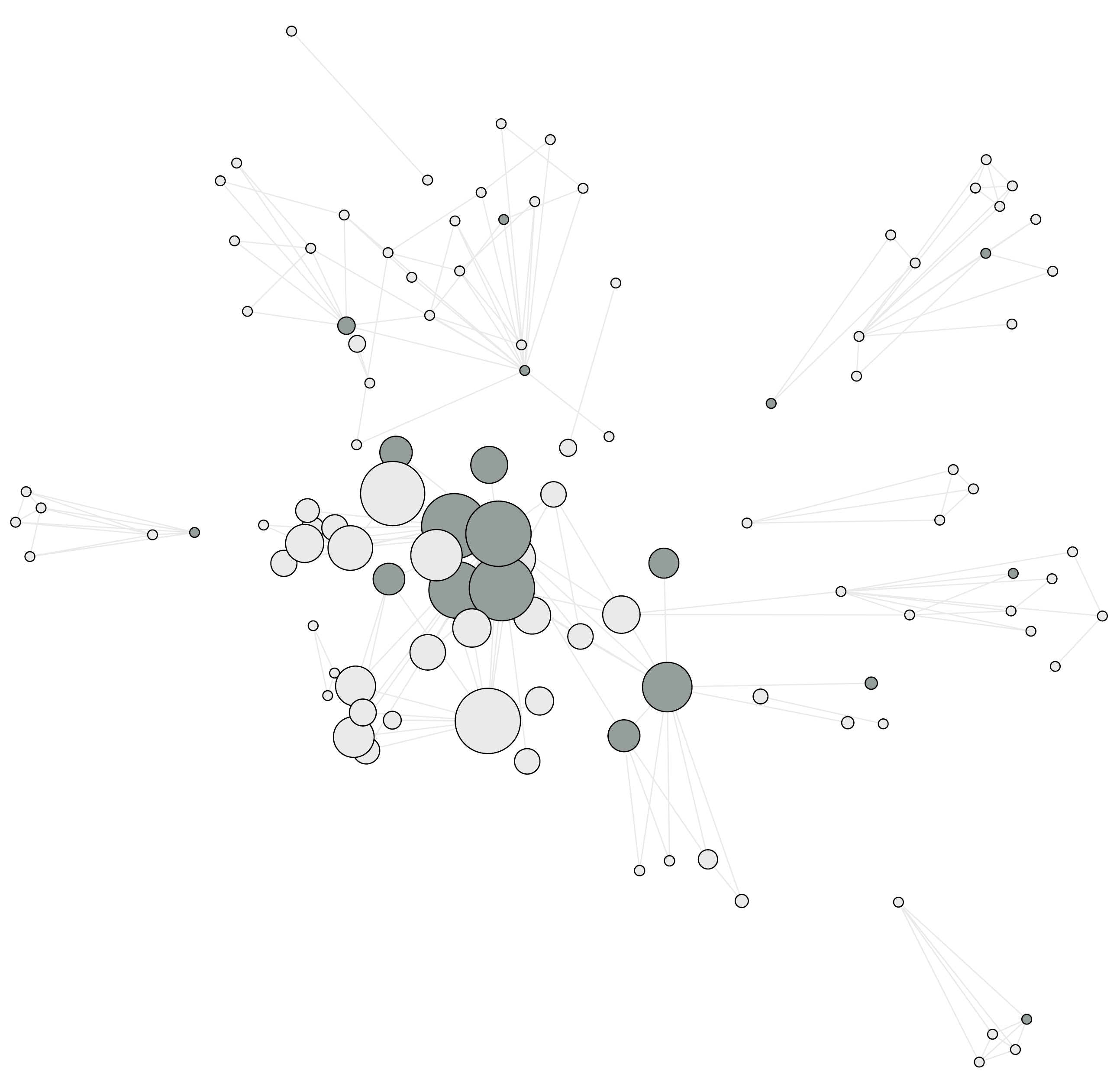}}
\subfigure[2004]{       \includegraphics[width=0.40\textwidth]{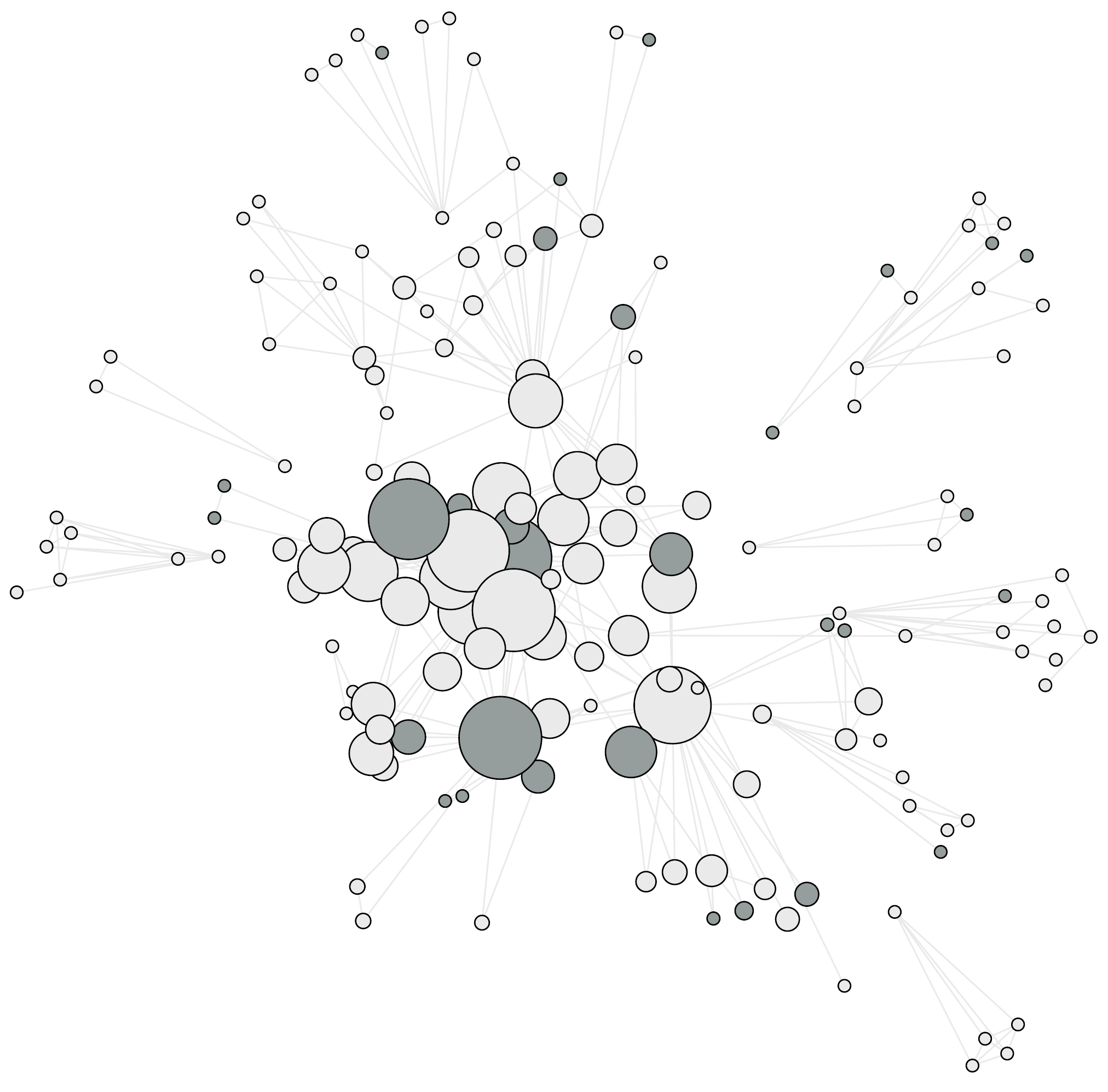}}
\subfigure[2005]{
	\includegraphics[width=0.40\textwidth]{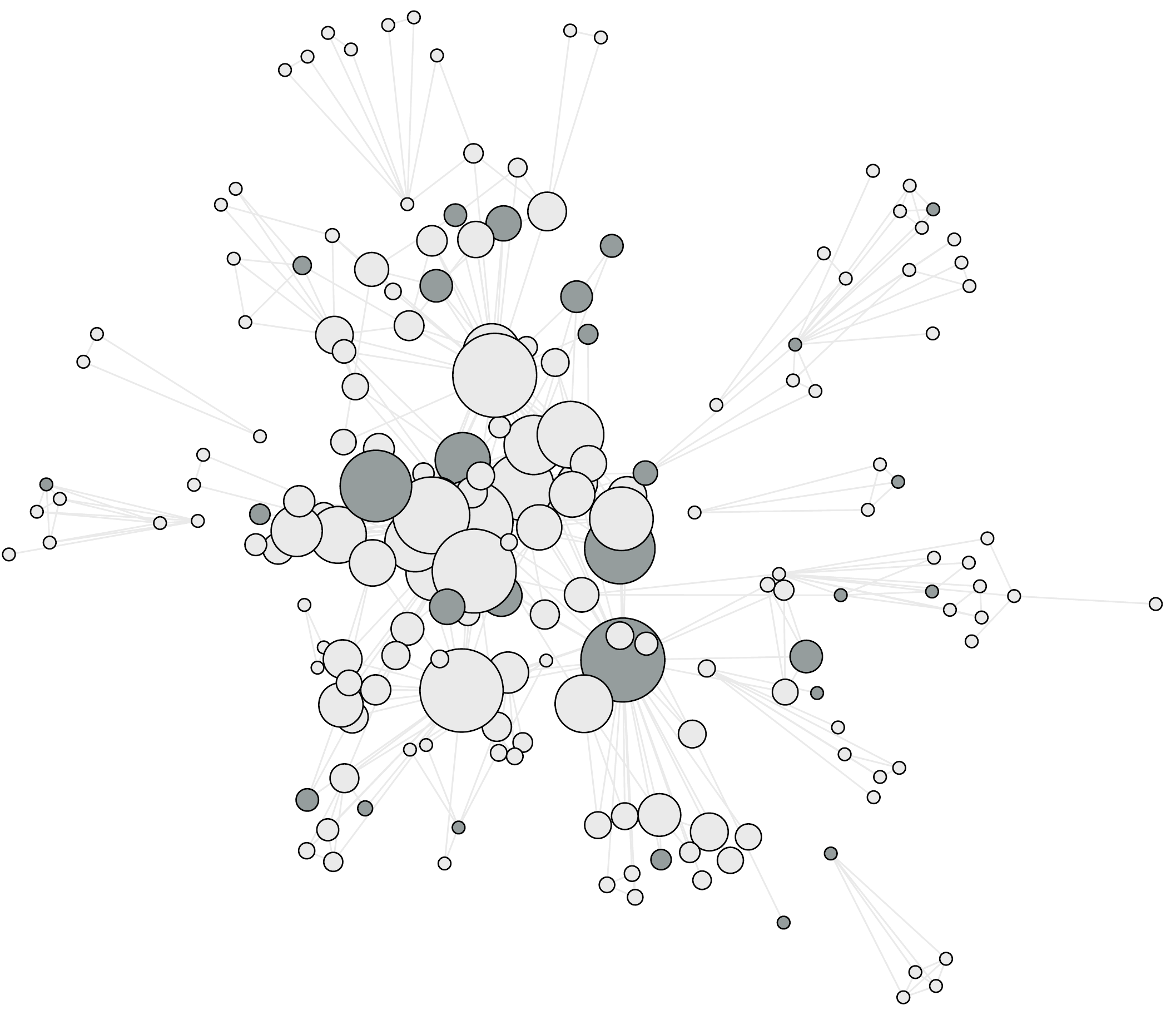}}
\subfigure[2006]{       \includegraphics[width=0.40\textwidth]{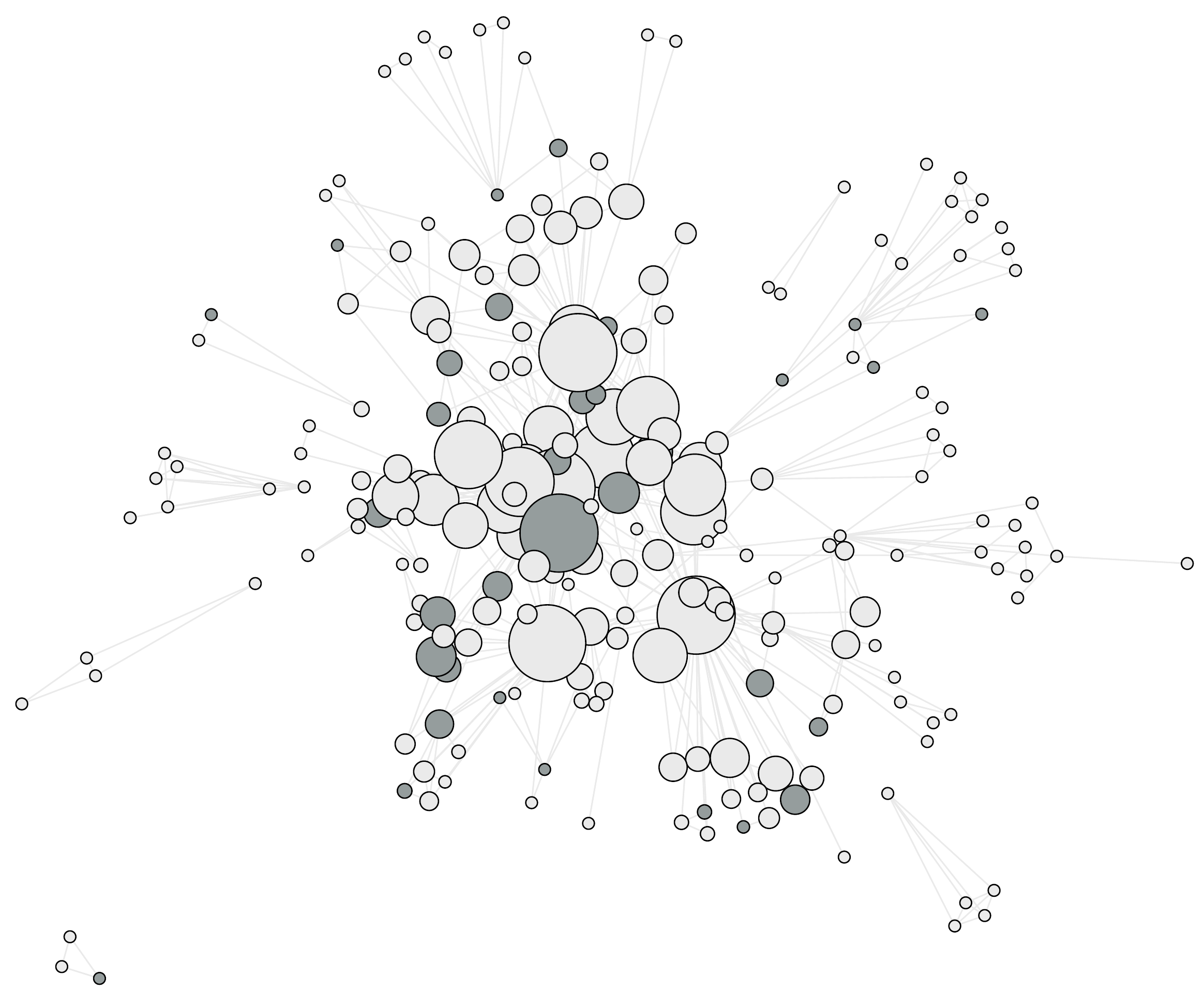}}
\subfigure[2007]{
	\includegraphics[width=0.40\textwidth]{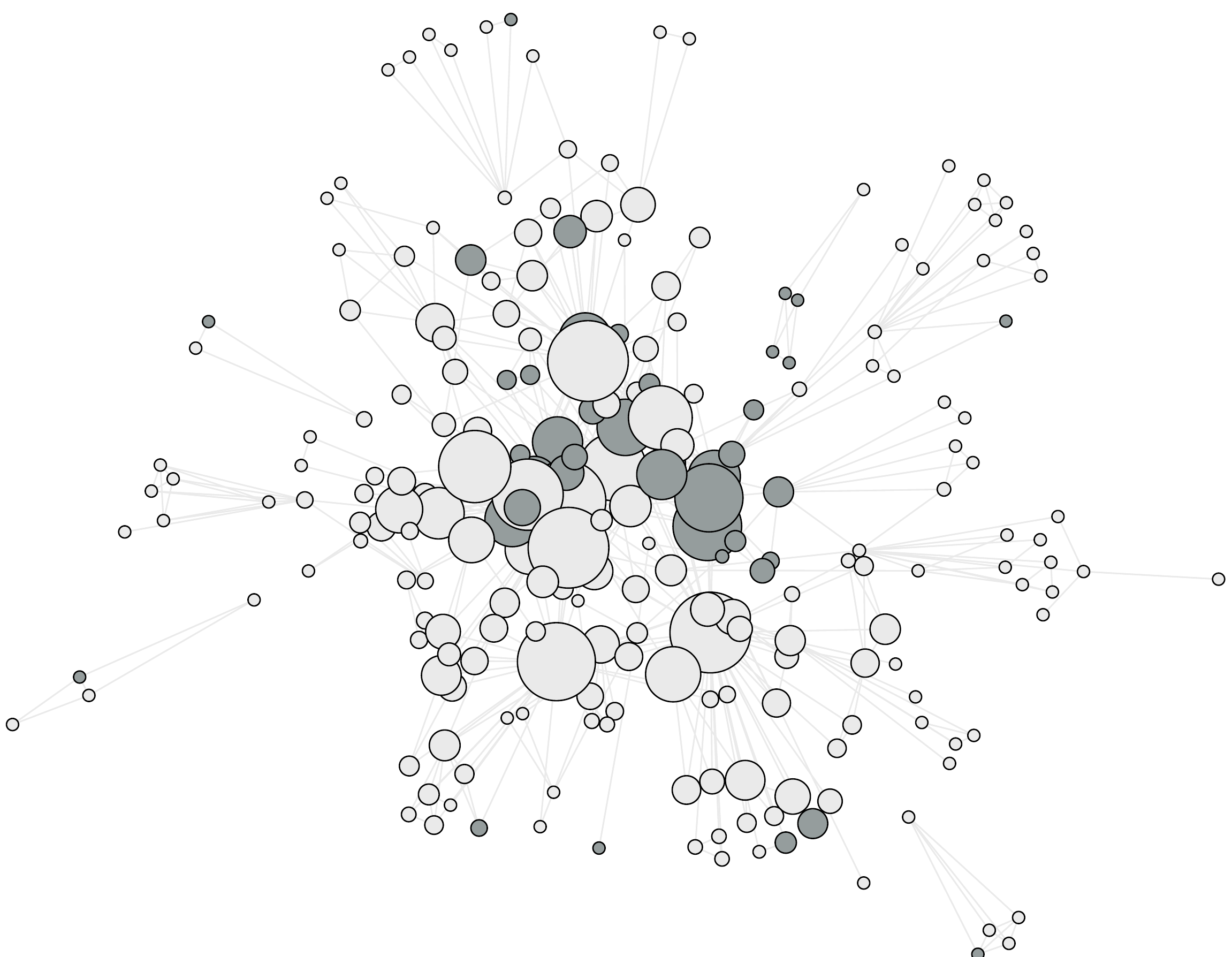}}
	\caption{\label{fig:networks}The CENS coauthorship network under study sliced according to year of publication (2002 through 2008, cumulative data). Node diameter represents the eigenvector centrality score of the node, where more central nodes have larger diameters. Moreover, the darker gray nodes denote individuals that are primarily affiliated with CENS.}
	\end{center}
\end{figure*}

The resulting network data were ``sliced'' according to publication year yielding to $7$ separate networks, each one representing the cumulative collaborative effort of CENS researchers up to that year. These networks of coauthorship for years from 2002 to 2007 are depicted in Figure \ref{fig:networks} (a through f).\footnote{The network depicting year 2001, though analyzed in this article, is not diagrammed as it is very sparse. Its visualization does not inform the present discussion.} In all depicted networks, the diameter of each node is defined by its eigenvector centrality \cite{power:bonacich1987} and the darker gray nodes denote individuals that are primarily affiliated with CENS.

\subsection{Preliminary analysis of network evolution: range, configuration, and topology}
For each one of the coauthorship networks under study, for years 2001 through 2007, we calculated some fundamental network statistics, presented in Table \ref{tab:netstats}. 

\begin{table*}[htp]
\centering
\begin{footnotesize}
\begin{tabular}{lllllllllll}
\hline
quantity/year&2001&2002&2003&2004&2005&2006&2007\\
\hline\hline
Authors (nodes)&35&68&127&203&228&278&291\\
Publications&23&69&175&303&418&496&547\\
Collaborations (edges)&182&346&690&1248&1598&2158&2536\\
\hline
Connected components&4&5&8&13&9&6&5\\
Diameter&3&4&6&7&8&8&7\\
\hline
Average path length, $\ell$&1.543&2.324&3.090&3.038&3.339&3.224&2.944\\
Clustering coefficient, $C$&0.645&0.543&0.432&0.387&0.389&0.337&0.329\\
Degree assortativity, $r$&0.432&0.299&0.272&0.187&0.180&0.166&0.165\\
\hline
\end{tabular}
\end{footnotesize}
\caption{\label{tab:netstats}A summary of the fundamental network statistics of the CENS coauthorship network for the time period 2001-2007: number of authors, publications and collaborations (range), number of connected components and diameter of the largest connected component (configuration), and average path length, clustering coefficient and degree assortativity (topology). All values presented are cumulative.}
\end{table*}

An analysis of these statistics provides insights into the evolution of the CENS coauthorship network over time. The first three rows of the table contain, for every year in the period under study, the cumulative number of a) authors, i.e. nodes in the network, b) publications (journal articles, conference papers, etc.), and c) collaborations (coauthoring events), i.e. edges in the network. When analyzed over time, these three values all follow a similar trend, which highlights two distinct time periods: a first term (2001-2004) during which all values increase sharply (roughly doubling in size from year to year), and a second term (2004-2007), during which the growth slows down. In particular the author count values indicate that CENS quickly became large and diversified in its population in the first term reaching a solid population base of collaborators by the year 2004. In the second term, from 2004 to 2007, the author population increased again, but to a much lesser extent (from 203 to 291 individuals), while the number of published works and collaborations maintained a regular growth (from 1248 to 2536 collaborations), suggesting the formation of a core CENS authoring base. 

This finding is confirmed by a quick analysis of the network's configuration. The number of connected components, i.e. the number of maximal connected subgraphs, goes from 4, in 2001, to 13, in 2004, indicating that the network becomes more fragmented in the first term, even if collaboration is overall increasing. In the second term, however, the number of connected components drops and the network quickly solidifies into a giant component, which indicates a solid base of strong collaboration. By looking at the the network diameter, i.e. the length of the longest geodesic path in the largest connected component, the formation of the giant component in year 2004 becomes evident. This is further reinforced by a quick analysis of Table \ref{tab:components}, which lists component populations by year.

\begin{table*}[htp]
\centering
\begin{footnotesize}
\begin{tabular}{lll}
\hline
year&\#&population\\
\hline\hline
2001&4&18 6 7 4\\
2002&5&38 16 8 4 2\\
2003&8&93 10 7 5 4 3 3 2\\
2004&13&150 10 8 8 5 4 3 3 3 3 2 2 2\\
2005&9&205 5 3 3 3 3 2 2 2\\
2006&6&262 5 4 3 2 2\\
2007&5&280 4 3 2 2\\
\hline
\end{tabular}
\end{footnotesize}
\caption{\label{tab:components}Node populations of connected components in the CENS coauthorship network, by year.}
\end{table*}

The preliminary analysis of these first two sets of values from Table \ref{tab:netstats} gives us a good understanding of the evolution of the range and configuration of the network over time. A third set of values, presented in Table \ref{tab:netstats} (average path length, clustering coefficient, and degree assortativity) can be investigated to provide an in depth understanding of its topology. 

The average path length is the average length of the shortest paths between all possible node pairs and, in turn, an indicator of the efficiency of information transfer in a social network \cite{wasserman_sna}. Short average path length, and thus high information transmission, are typical characteristics of many real and small-world networks \cite{watts_nature_98}. In the network under study, the average path length is about $1.5$ in 2001; it grows steadily in the first term, reaching a value of about $3.0$, which stays roughly constant throughout the second term. This indicates that once the CENS authoring base is formed, an average of $3$ steps are necessary to transfer information among any two pairs of nodes. Remarkably, this value resembles more closely that found in typical small-world networks, such as movie actors ($\ell = 3.65$) \cite{watts_nature_98}, than that of scholarly coauthorship networks, such as mathematics ($\ell = 9.5$) and neuroscience ($\ell = 6$) \cite{barabasi2002}; yet, this relatively low average path is due to the relatively small size of the network analyzed (the mathematics and neuroscience coauthorship networks have $N = 70,975$ and $209,293$, respectively). 

The clustering coefficient measures the density of clique-like triangles in a network. High clustering coefficient coupled with short average path length indicates that a network exhibits small-world properties \cite{newman-siam}. In the CENS coauthorship network, the clustering coefficient decreases steadily over time, from an initial value of $0.645$ in 2001, to $0.329$ in 2007. This suggests that the network becomes less cliquish and collaboration patterns becomes more uniform across the network over time. This trend reveals that the CENS network initially matches the typical topology of highly-clustered disciplines such as physics ($C = 0.56$) and biology ($C = 0.6$) \cite{newman-siam} but later drops to the values normally recorded in less cliquish domains, such as mathematics ($C = 0.34$) \cite{Grossman95}.

A final indicator of network topology presented here is degree assortativity. In general, assortativity can be defined as the tendency for individuals (nodes) in a social network to establish connections preferentially to other individuals with similar characteristics \cite{newman:assort}. The most common measure of assortativity is computed based on the individuals' degree centrality. In the network presented here, degree-based assortativity decreases steadily over time from a value of $0.432$ in 2001 to $0.165$ in 2007. Interestingly, the decline of the degree assortativity measure follows very closely that of the clustering coefficient --- the correlation between the two is $\rho = 0.964$ (p-value $< 0.005$). This means that as collaboration patterns in the network become more sparse and uniform (decreasing $C$), they also become more mixed (decreasing $r$), i.e. highly-connected individuals begin to collaborate with lowly-connected ones. In the next section, we extend the study of assortativity to a set of discrete characteristics, namely authors' academic department, affiliation, position, and country of origin. Analysis of these mixing patterns allows us to understand the homophilious and heterophilious components that contribute to the observed growth of the network.

\section{Studying network evolution in terms of discrete node characteristics}

The preliminary analysis of the CENS coauthorship network presented in the previous section reveals the following scenario. In 2001, the network of collaboration is small and very fragmented. During the first few years of activity, however, the CENS group grows significantly in the number of authors and collaborations. By the end of 2004, a solid base of collaborating authors (i.e.~a giant component) is formed. In the analyzed network, small-world effects become less prominent over time; in particular, average distance between individuals becomes larger (increasing $\ell$), and collaboration patterns in the network become more sparse (decreasing $C$) and more mixed (decreasing $r$).  

Although our preliminary analysis presents a fairly comprehensive account of the range, configuration and topology of the studied network of scientific collaboration over time, we believe that it fails to provide a sociological explanation of the dynamics underlying the observed patterns. In particular, we were curious to explore further the correlation between clustering coefficient and assortativity. Our preliminary analysis indicates that there exists a solid link between these two patterns: a) the network becoming more sparse and uniform, and b) collaboration patterns becoming more mixed. However, this analysis is restricted to degree assortativity and thus ignores other mixing patterns that might have contributed to the decrease in network clustering over time. For this reason, we became interested in deepening our understanding of the sociological and academic context of the CENS collaboration network to identify specific patterns that might account for the observed clustering pattern. For example, can we speculate that the network becoming more sparse is indicative of higher interdisciplinary collaboration and/or higher collaboration across different institutions? In this context, the question that we would like to address is: what specific mixing patterns are accountable for the decrease in the network's clustering coefficient? In the remainder of this article, we extend the temporal analysis, presented in the previous section, to a set of node characteristics. Our aim is to justify the observed clustering pattern in terms of specific social and academic characteristics that are more telling of the sociological aspects of the group than what degree assortativity alone can elucidate.

\subsection{Further data collection}
We collected these additional metadata relative to each author in the network under study: a) academic affiliation, b) academic department, c) academic position, and d) country of origin.

\begin{table*}[htp]
\centering
\begin{footnotesize}
\begin{tabular}{ll}
\hline
\textbf{Academic affiliation}&\\
\hline\hline
University of California, Los Angeles (UCLA)&148\\
University of Southern California (USC)&66\\
Massachusetts Institute of Technology (MIT)&10\\
California Institute of Technology (Caltech)&8\\
University of California, Riverside (UC Riverside)&7\\
University of California, Berkeley (UC Berkeley)&7\\
University of California, Merced (UC Merced)&4\\
State University of New York at Stony Brook (SUNYSB)&3\vspace*{0.1in}\\

\hline
\textbf{Academic department}&\\
\hline\hline
Computer Science&113\\
Electrical Engineering&80\\
Civil Engineering&23\\
Biology&19\\
Information Sciences&9\\
Environmental Science&7\\
Education&5\\
Marine Biology&4\\
Statistics&3\\
Linguistics&3\vspace*{0.1in}\\

\hline
\textbf{Academic position}&\\
\hline\hline
Graduate Student&97\\
Staff / Research Associate (Staff)&67\\
Full Professor (Professor)&44\\
Postdoctoral Student (PostDoc)&21\\
Associate Professor&21\\
Assistant Professor&20\\
Undergraduate Student&5\\
Lecturer&3\vspace*{0.1in}\\

\hline
\textbf{Country of origin}&\\
\hline\hline
United States of America (USA)&120\\
India&33\\
China&24\\
Italy&10\\
South Korea (Korea)&9\\
Australia&5\\
Greece&4\\
Iran&3\\
Taiwan&3\\
Mexico&3\\

\end{tabular}
\end{footnotesize}
\caption{\label{tab:pop}Population counts (10 most recurring values) for the collected node properties: authors' academic affiliation, department, position and country of origin}
\end{table*}

We collected these metadata via manual techniques, i.e.~gathering required information on the authors' personal web pages and consulting online directories from university and department web sites. It is worth noting that all the parameters collected (except for country of origin) are subject to change over time, even in the short timespan studied in this article. Researchers and scientists might change institution, department and position in a seven-year time period. For this reason, we consulted not only authors' personal web sites, but also their curriculum vitae and biographies to record changes in their academic affiliation, department and position. Curriculum vitae were also useful to collect authors' country of origin, which, for the purpose of this study, we consider to be the country in which individuals pursued their high-school education. Table \ref{tab:pop} presents the frequency counts for the collected author metadata: authors' academic affiliation, department, position and country of origin.

A quick analysis of Table \ref{tab:pop} reveals that the collaboration network studied here consists mostly of scholars from UCLA and USC from the departments of Computer Science and Electrical Engineering. The network consists of a large number of graduate students,\footnote{Graduate students are students both at Ph.D. and Master level. Although these have been incorporated into a single category, Ph.D. students are the vast majority in this group (over 95\%).} but researchers and professors at all levels make up a large portion of the collaboration network. Finally, the network is very international in its population, with about half of the authors being from the United States, about a quarter from India and China and the rest from a number of other countries worldwide.

\subsection{Analysis of network evolution by discrete assortative mixing}

The temporal analysis of degree assortativity, presented in the previous section, indicates the extent to which individuals in the network co-author preferentially to other individuals with similar degree centrality. Using the newly collected author metadata --- academic affiliation, department, position and country of origin --- we can extend our investigation of assortativity to compute mixing patterns based on these discrete parameters. In our case, all analyzed parameters are nominal and we can thus measure discrete assortativity coefficient, $r$, using the following formula \cite{newman:mixpatt2003}:
\begin{equation}
r = \frac{\sum_i e_{ii} - \sum_i a_i b_i}{1-\sum_i a_i b_i}
\end{equation}
where $e_{ij}$ is the fraction of edges in a network that connect a node of type~$i$ to one of type~$j$, $a_i$ is the fraction of edges that have a node of type $i$ on the head of the edge, and $b_i$ is the fraction of edges that have a node of type $i$ on the tail of the edge. Finally, $r = -1$ when there is perfect disassortative mixing, $r = 0$ when there is no assortative mixing, and $r = 1$ when there is perfect assortative mixing. In other words, the discrete assortativity coefficient, $r$, indicates the level of homophily of the network for a certain parameter. For example, if $r$ for academic affiliation is 1.0, this means that individuals in the network only write papers with other individuals with same institutional affiliation. In this kind of network, there are no multi-institutional collaborations. On the other side of the spectrum, we can imagine a completely disassortative network ($r = -1$) in which every single collaboration (i.e. paper) in the network is authored by individuals that belong to different institutions. 

\begin{figure*}[h]
\centering
\includegraphics[width=1.0\textwidth]{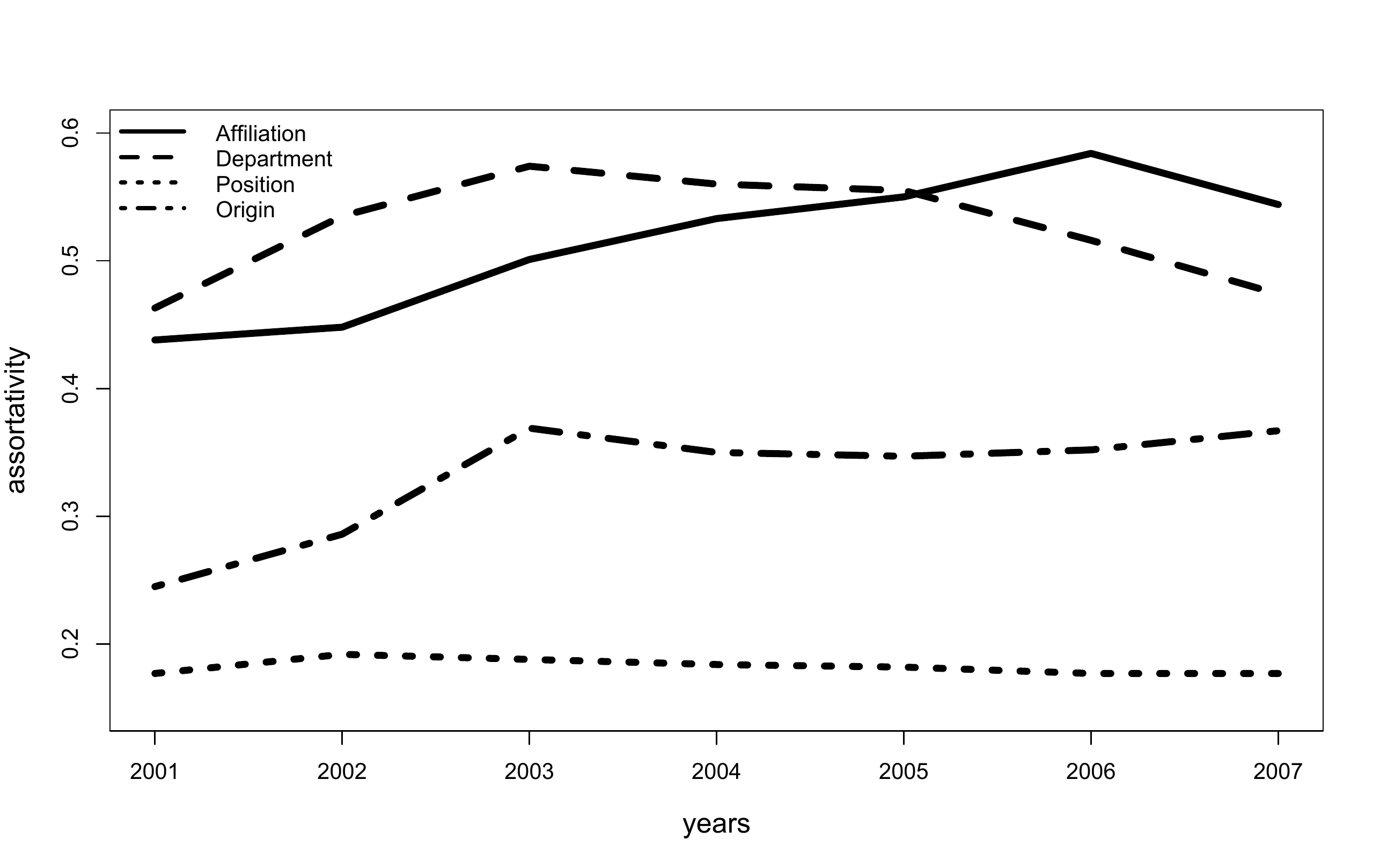}
\caption{\label{fig:assort}Plot of discrete assortativity coefficients for years 2001 through 2007 based on authors' academic affiliation, department, position and country of origin}
\end{figure*}

Table \ref{tab:assort} presents the discrete assortativity coefficients for the network under study based on authors' academic affiliation, department, position and country of origin, calculated at seven temporal snapshots of the network (2001 through 2007). A visual representation of these values is also presented in the plot of Figure \ref{fig:assort}.

\begin{table*}[h]
\centering
\begin{footnotesize}
\begin{tabular}{lllllllllll}
\hline
property/year&2001&2002&2003&2004&2005&2006&2007\\
\hline\hline
Academic affiliation&0.438&0.448&0.501&0.533&0.550&0.584&0.544\\
Academic department&0.463&0.535&0.574&0.560&0.555&0.516&0.474\\
Academic position&0.177&0.192&0.188&0.184&0.182&0.177&0.177\\
Country of origin&0.245&0.286&0.369&0.350&0.347&0.352&0.367\\
\hline
\end{tabular}
\end{footnotesize}
\caption{\label{tab:assort}Discrete assortativity coefficients for years 2001 through 2007 based on authors' academic affiliation, department, position and country of origin}
\end{table*}

Looking at Figure \ref{fig:assort}, it is easy to deduce that these assortativity coefficients are all within the same range --- between a minimum of $0.1$ and a maximum of $0.6$. Also, they do not change very much over the period under study --- fluctuations in the 7-year period rarely exceed $0.1$. Overall, the network is more assortative by academic affiliation and department, and less assortative by academic position and country of origin. A detailed temporal analysis of these coefficients can provide insights into the extent and nature of mixing patterns in the coauthorship network. For example, by analyzing the assortativity by academic department, we can deduce how inter-disciplinary the CENS network is and how inter-disciplinarity has changed over time. However, this analysis fails to reveal the specific components that contributed to these mixing patterns, e.g. \textit{what collaborations are most responsible for the increase in inter-disciplinarity?}. In the remainder of this section we present, for each one of the studied characteristics, a detailed interpretation of our findings. Our aim is to push our understanding of the assortativity coefficients further, decomposing the observed collaboration patterns along specific components, to allow a more in-depth temporal analysis of the observed mixing patterns.
 
\subsubsection{Academic affiliation}
From Figure \ref{fig:assort}, the assortativity coefficient based on nodes' academic affiliation grows steadily over time, by about $0.1$, from $0.438$ in 2001 to $0.544$ in 2007. This indicates that, overall, authors in the CENS coauthorship network have increasingly collaborated with other authors from the same institutional affiliation, in the time period under study. In the latest snapshot of the CENS network (year 2007) academic affiliation is the single most assortative characteristic, suggesting that CENS authors collaborate preferentially with individuals in their institution. This finding matches an earlier observation that the community structure of CENS collaboration matches very closely its institutional configuration \cite{rodrpepe}. 

\begin{figure*}[h]
\centering
\includegraphics[width=1.0\textwidth]{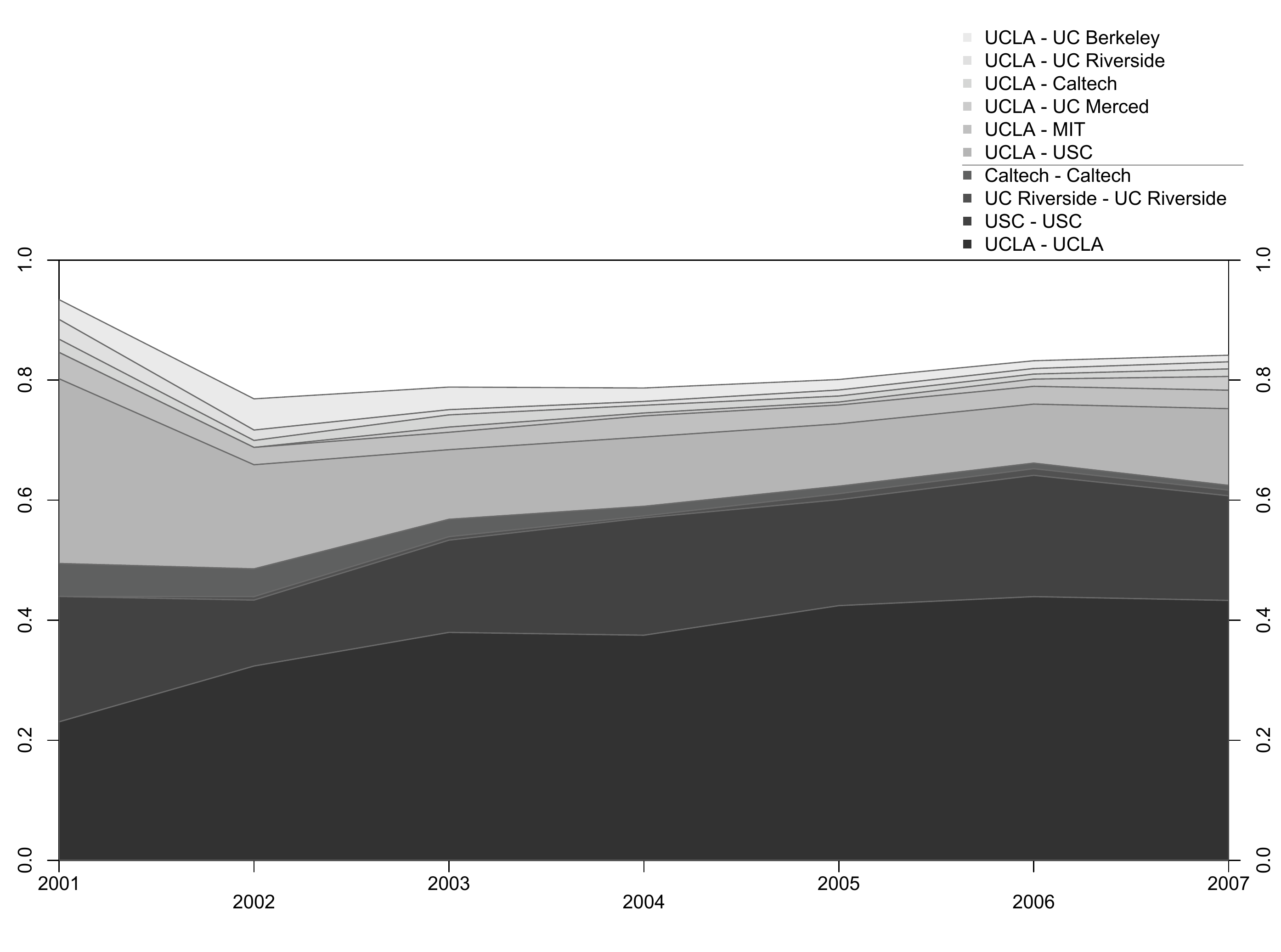}
\caption{\label{fig:affil_stack} Top ten most recurrent academic affiliation pairs as fraction of total volume of collaboration. Darker polygons at the bottom are intra-institutional collaboration, while lighter polygons depict inter-institutional collaboration.}
\end{figure*}

We would like to investigate this finding further, analyzing the specific intra- and inter-institutional collaborations that contributed to making the network more assortative over time. In order to do this, we inspect the most recurrent mixing patterns by affiliation in the last snapshot of the network, i.e., the institutional pairs that make up the majority of collaboration volume in 2007. These values are presented in Table \ref{tab:affil}. Clearly, as the coauthorship network under study is undirected, the order of the pairs (affil-1, affil-2) is not relevant. The top four rows in Table \ref{tab:affil} present the institutional pairs contributing to intra-institutional collaboration, whereas the bottom six rows present the pairs contributing to inter-institutional collaboration. So for example, in year 2001 there were 42 coauthorship activities among individuals affiliated with UCLA. From a network perspective, this means that in the year 2001, 42 edges, out of a total of 182 (from Table \ref{tab:netstats}), were among nodes with parameter ``UCLA''. Collaboration between UCLA researchers increases steadily and reaches 1098 (out of a total of 2536) edges in 2007. In Figure \ref{fig:affil_stack}, we plot the values of Table \ref{tab:affil} as a fraction of the total volume of collaborations each year. Please note that the order of the institution pairs in Figure \ref{fig:affil_stack} is reversed with respect to Table \ref{tab:affil}, so that most prominent collaborations occupy the lower portion of the plot. 

\begin{table*}[htp]
\centering
\begin{footnotesize}
\begin{tabular}{lllllllll}
\hline
\textbf{affil-1}&\textbf{affil-2}&2001&2002&2003&2004&2005&2006&2007\\
\hline\hline
UCLA&UCLA&42&112&262&468&678&948&1098\\
USC&USC&38&38&106&244&282&436&442\\
UC Riverside&UC Riverside&-&2&4&4&16&24&24\\
Caltech&Caltech&10&16&20&20&20&20&20\\
\hline
UCLA&USC&56&60&80&144&166&212&324\\
UCLA&MIT&8&10&20&44&50&64&78\\
UCLA&UC Merced&-&-&6&6&8&26&58\\
UCLA&Caltech&4&4&14&16&16&18&32\\
UCLA&UC Riverside&6&6&6&8&16&20&30\\
UCLA&UC Berkeley&6&18&26&28&28&28&28\\
\hline
\end{tabular}
\end{footnotesize}
\caption{\label{tab:affil} Most recurrent academic affiliation pairs (top 10 results, cumulative values). Top 4 rows show intra-institutional collaboration. Bottom 6 rows show inter-institutional collaboration.}
\end{table*}

The stacked plot of Figure \ref{fig:affil_stack} allows us to decompose the assortativity coefficient trend lines of Figure \ref{fig:assort} for discrete parameter academic affiliation. From Figure \ref{fig:assort}, assortativity by affiliation increases steadily from 2001 to 2006 and finally drops slightly from 2006 to 2007. This trend can be understood in terms of the growth of intra- and inter-institutional collaborations, presented in Figure \ref{fig:affil_stack}. From the stacked plot of Figure \ref{fig:affil_stack}, we note that in 2001, the vast majority of recorded collaborations involves intra- and inter-institutional efforts between UCLA and USC individuals. In year 2007, the picture is not very different, with UCLA and USC still composing the bulk of the total volume of collaborations. However, a closer look at the components of the plot reveals that intra-institutional collaboration at UCLA has doubled in volume (from 0.2 to 0.4) while inter-institutional collaboration (UCLA-USC) has halved (0.3 to 0.15), compared to 2001 values. USC-USC collaboration stays roughly constant throughout the period under study. The increase in UCLA-UCLA and the decrease of UCLA-USC collaborations are the components that are most responsible for the increase in assortativity coefficient by affiliation from 2001 to 2007, presented in Figure \ref{fig:assort}. 

There are some other collaboration dynamics that contribute to this trend. For example, besides UCLA-USC, the overall inter-institutional effort of UCLA decreases (e.g. collaborations with UC Berkeley and UC Riverside). Moreover, intra-institutional collaborations by Caltech researchers (which make almost 10\% of the total volume in 2001) fade away over time. In sum, by year 2007, the collaboration scenario at CENS is largely dominated by publications authored within UCLA. Based on this finding, we can conclude that despite CENS's mission to be a multi-institutional research center, the temporal decomposition of coauthorship patterns demonstrates that CENS collaboration became less inter-institutional from 2001 to 2007 consolidating around its main institution, UCLA. The steady increase in UCLA-UCLA collaboration can possibly be attributed to the construction of a CENS headquarter office at UCLA, completed in 2005. We can speculate that the CENS headquarter has brought UCLA scholars closer to each other, enabling interpersonal collaboration among them and, in turn, boosting coauthorship activity.\footnote{This is a fair assumption especially for the authoring of scientific conference papers, that have a much quicker publication turnaround than journal articles.}

\subsubsection{Academic department}
\begin{figure*}[h!]
\centering
\includegraphics[width=1.0\textwidth]{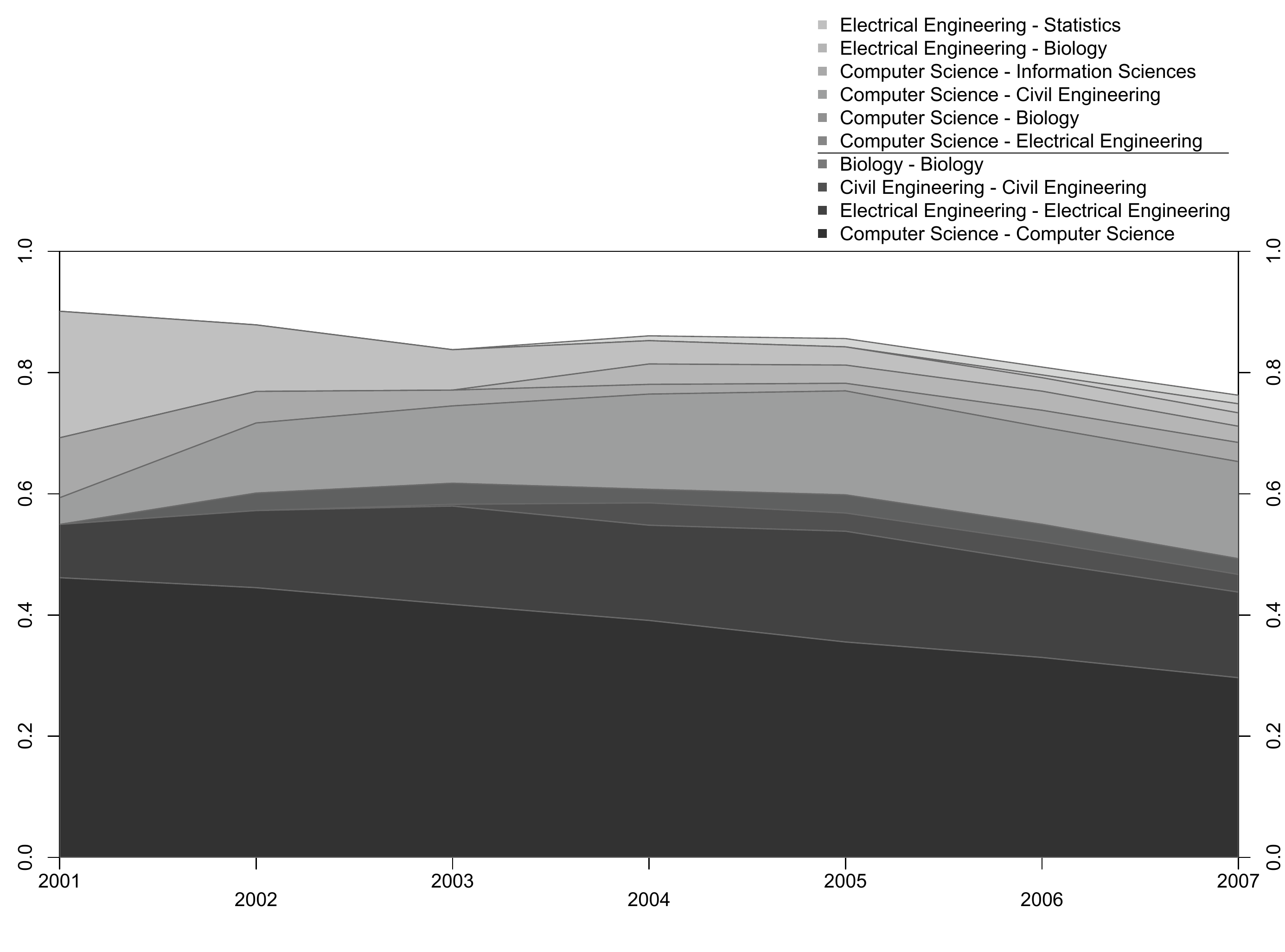}
\caption{\label{fig:dep_stack} Top ten most recurrent academic department pairs as fraction of total volume of collaboration. Darker polygons at the bottom are intra-departmental collaboration, while lighter polygons depict inter-departmental collaboration.}
\end{figure*}

From Figure \ref{fig:assort}, the assortativity coefficient for academic department has the following trend. In year 2001, the CENS network is heavily assortative based on academic department ($r = 0.463$). In the following two years, assortativity increases even more, reaching a peak of $0.574$ in 2003. This means that in 2003, the CENS coauthorship was very highly fragmented by department. By extension, we can speculate that at this time, collaboration patterns were vastly mono-disciplinary. In later years, however, assortativity by department decreases. Even though the value recorded in 2007 ($r = 0.474$) is roughly equivalent to the network's outset, the trend observed from 2003 to 2007 indicates the CENS collaboration network becoming more interdisciplinary. A decomposition of the observed coauthorship patterns, presented below, enables us to understand, more in depth, the extent of intra- and inter-departmental collaboration. We present in Table \ref{tab:dept} the most recurrent mixing pattern pairs by academic department. The stacked plot of Figure \ref{fig:dep_stack} depicts these values as fraction of total volume of collaboration.

\begin{table*}[h]
\centering
\begin{footnotesize}
\begin{tabular}{lllllllll}
\hline
\textbf{dept-1}&\textbf{dept-2}&2001&2002&2003&2004&2005&2006&2007\\
\hline\hline
Computer Sci&Computer Sci&84&154&288&488&568&712&752\\
Electrical Eng&Electrical Eng&16&44&112&196&292&338&358\\
Civil Eng&Civil Eng&-&-&2&46&48&74&74\\
Biology&Biology&-&10&24&28&48&62&66\\
\hline
Computer Sci&Electrical Eng&8&40&88&196&274&346&406\\
Computer Sci&Biology&18&18&18&20&20&60&80\\
Computer Sci&Civil Eng&-&-&-&42&48&68&68\\
Computer Sci&Information Sci&38&38&46&48&48&48&56\\
Electrical Eng&Biology&-&-&-&-&-&10&38\\
Electrical Eng&Statistics&-&-&-&10&22&28&36\\
\hline
\end{tabular}
\end{footnotesize}
\caption{\label{tab:dept} Most recurrent academic department pairs (top 10 results, cumulative values). Top 4 rows show intra-departmental collaboration. Bottom 6 rows show inter-departmental collaboration.}
\end{table*}

In Figure \ref{fig:dep_stack}, the darker, bottom four polygons are the department pairs that contribute the most to intra-departmental collaboration, whereas the top six polygons present the department pairs that account for most of inter-departmental collaboration. At the network's outset, CENS collaboration is dominated by intra-departmental collaborations in Computer Science and Electrical Engineering. The most prominent inter-departmental collaborations are between Computer Science and both Biology and Information Sciences. The increase in the assortativity coefficient by department from 2001 to 2003 (shown in Figure \ref{fig:assort}) can be attributed to a number of factors, including (i) a slight increase in collaborations among Electrical Engineers, (ii) the appearance of novel collaborations among Biologists, and (iii) a substantial drop in collaborations by Computer Scientists with both Biologists and Information Scientists.

In the long run, however, the intra-departmental volume of collaboration among Computer Scientists decreases steadily over time. This decrease, coupled with the growth of a number of inter-departmental collaborations (Computer Science with Electrical and Civil Engineering, as well as Electrical Engineering with Biology and Statistics), is most responsible for the assortativity coefficient trend presented in Figure \ref{fig:assort}, i.e., the CENS coauthorship network becomes less assortative by department, and thus more inter-disciplinary, over time. 

The observed patterns can be interpreted as follows. First the overall presence of intra-departmental collaborations in Computer Science is telling of the nature of research being performed at CENS. The domain of networked sensing emerges historically from computer network research and is thus, normally located as a branch in departments of Computer Science. Sensor network technologies, however, require the design and construction of wireless sensors, and, in turn, interaction between computer sciences and engineering disciplines follows necessarily. This growing incidence of a core set of Electrical Engineering collaboration (both intra- and inter-departmental) is evident in Figure \ref{fig:dep_stack}. It is interesting to note that inter-departmental collaborations with Electrical Engineers involve not only Computer Science, but also Biology (a major scientific application area for sensor networks) and Statistics (a discipline increasingly required by field scientists to deal with issues related to sensor data cleaning, analysis, and modeling). Finally, it is worth noting that the volume of intra- and inter-departmental collaborations involving the department of Civil Engineering increases over time, possibly reflecting the inception in 2004 of a new application area at CENS involved in the development and application of sensing technologies in urban and social settings. 

In sum, CENS, a research center emerged as a sub-domain of Computer Science, has progressively become more inter-disciplinary over time. The increase in inter-disciplinarity can be attributed to CENS' need to develop sensor network technologies (Electrical Engineering), apply and deploy them in field environments (Biology and Civil Engineering), and concurrently deal with data analysis issues (Statistics).

\subsubsection{Academic position}
\begin{figure*}[h!]
\centering
\includegraphics[width=1.0\textwidth]{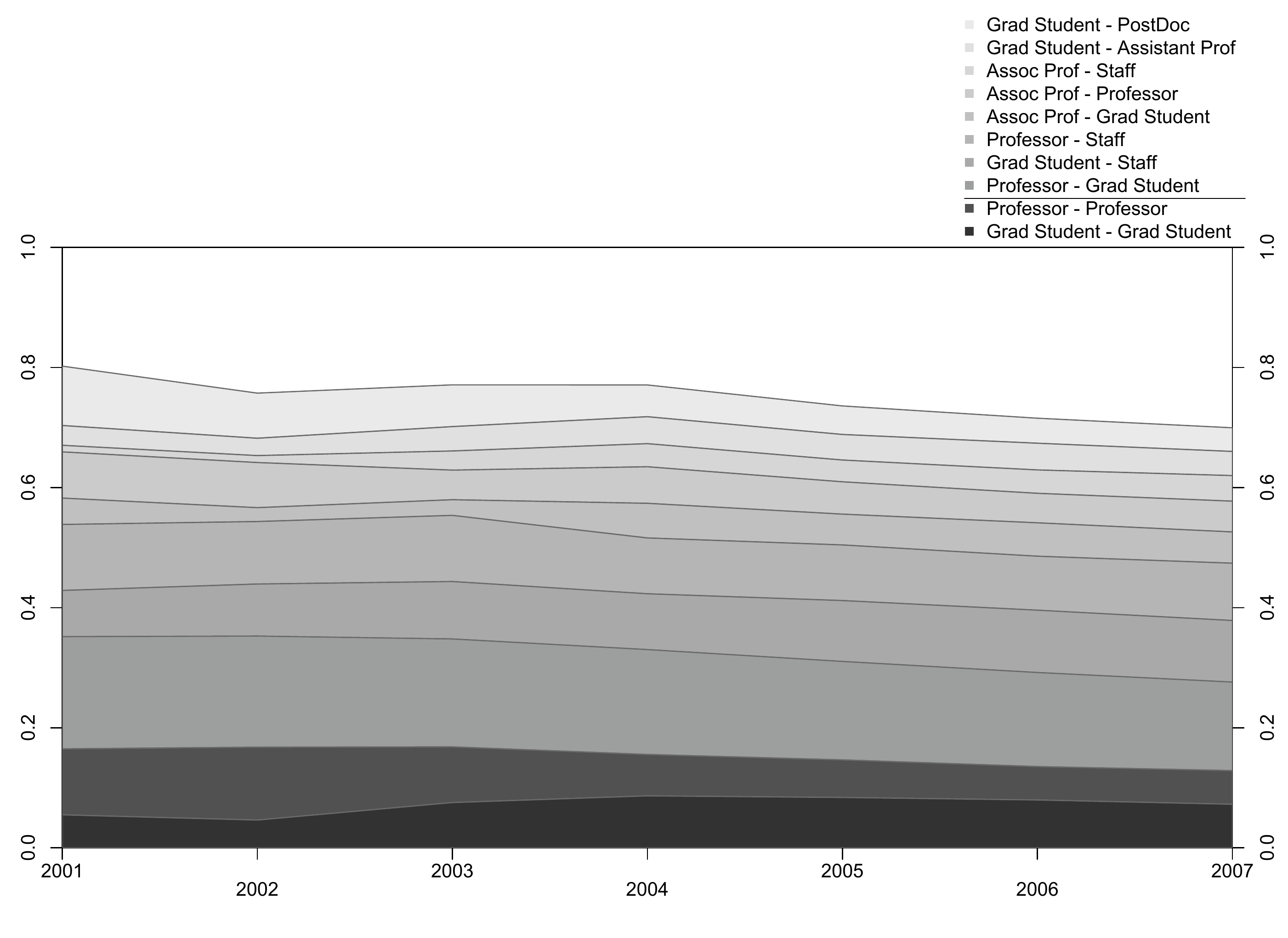}
\caption{\label{fig:pos_stack} Top ten most recurrent academic position pairs as fraction of total volume of collaboration. Darker polygons at the bottom depict collaborations among pairs of individuals with the same academic position, while lighter polygons depict collaborations between individuals of different academic positions.}
\end{figure*}

From Figure \ref{fig:assort}, it is evident that the discrete assortativity coefficient based on academic position has very little influence on the overall topology of the CENS coauthorship network, compared to the other computed measures. Assortativity by academic position never exceeds a value $0.2$ and, very importantly, it remains practically unchanged throughout the period under study. This finding suggests that the network is very weakly assortative with respect to academic position, i.e. coauthorship activities involve the collaboration among scholars of all ranks. It is fair to speculate that this finding is a \textit{de facto} characteristic of scholarly publishing in many academic fields. However, a detailed decomposition of the most prominent academic position pairs can help reveal the specific mixing patterns that influence this assortativity value of the CENS coauthorship network. Table \ref{tab:pos} presents the top ten most prominent academic position pairs and the stacked plot of Figure \ref{fig:pos_stack} presents these values as fraction of the cumulative, yearly volume of collaboration in the CENS network. 

\begin{table*}[h]
\centering
\begin{footnotesize}
\begin{tabular}{lllllllll}
\hline
\textbf{position-1}&\textbf{position-2}&2001&2002&2003&2004&2005&2006&2007\\
\hline\hline
Graduate&Graduate&10&16&52&108&134&172&184\\
Professor&Professor&20&42&64&86&100&120&142\\
\hline
Professor&Graduate&34&64&124&218&262&338&374\\
Staff&Graduate&14&30&66&116&162&224&260\\
Professor&Staff&20&36&76&116&148&194&242\\
Associate Prof&Graduate&8&8&18&72&82&120&132\\
Associate Prof&Professor&14&26&34&76&86&106&130\\
Associate Prof&Staff&2&4&22&48&58&84&108\\
Assistant Prof&Graduate&6&10&28&56&68&96&102\\
PostDoc&Graduate&18&26&48&66&76&90&100\\
\hline
\end{tabular}
\end{footnotesize}
\caption{\label{tab:pos} Most recurrent academic position pairs (top 10 results, cumulative values). Top 2 rows show collaboration between individuals with same position. Bottom 8 rows show collaboration between individuals in different positions.}
\end{table*}

In Figure \ref{fig:pos_stack}, the two darker polygons at the bottom depict the position pairs that contribute the most to make the network more assortative --- collaborations between both graduate students and between full professors. The eight polygons in lighter shades of gray depict the volume of collaborations that make the network less assortative, for they represent coauthorship activity among scholars of different ranks. 

The plot of Figure \ref{fig:pos_stack} is very variegated and none of the analyzed pairs stands out for volume of collaboration. Moreover, the overall assortativity coefficient by academic position remains constant over time. Yet, certain academic position pairs display some minor fluctuations which we deem interesting. Looking at the darker polygons in Figure \ref{fig:pos_stack} we note that collaboration among full professors diminishes slightly from 2001 to 2007, while coauthorship between graduate students increases, in the same period. This result suggests the following research scenario: at its outset, CENS did not include many graduate students; most of the research to ``bootstrap'' the research center was carried out by faculty members. However, as the center grew in size, more and more graduate students became involved and patterns of collaboration among graduates became more prominent. 

This finding is also reflected in the analysis of collaborating pairs of different ranks, i.e., the lighter polygons in the top portion of Figure \ref{fig:pos_stack}. Again, no major fluctuations emerge from a quick visual analysis of the plot. However, if we look collectively at collaborations of full professors with graduate students, staff members and other professor ranks, we note that they decrease slightly from 2001 to 2007, while collaborations among graduate students, staff, associate and assistant professors increase during the same time period. This finding reinforces the research scenario presented above; it indicates that after a ``bootstrap'' phase in which scholarly publication was led by high-ranking professors, a younger research network was formed, consisting of graduate students and faculty in the early stage of their careers.

\subsubsection{Country of origin}
\begin{figure*}[h!]
\centering
\includegraphics[width=1.0\textwidth]{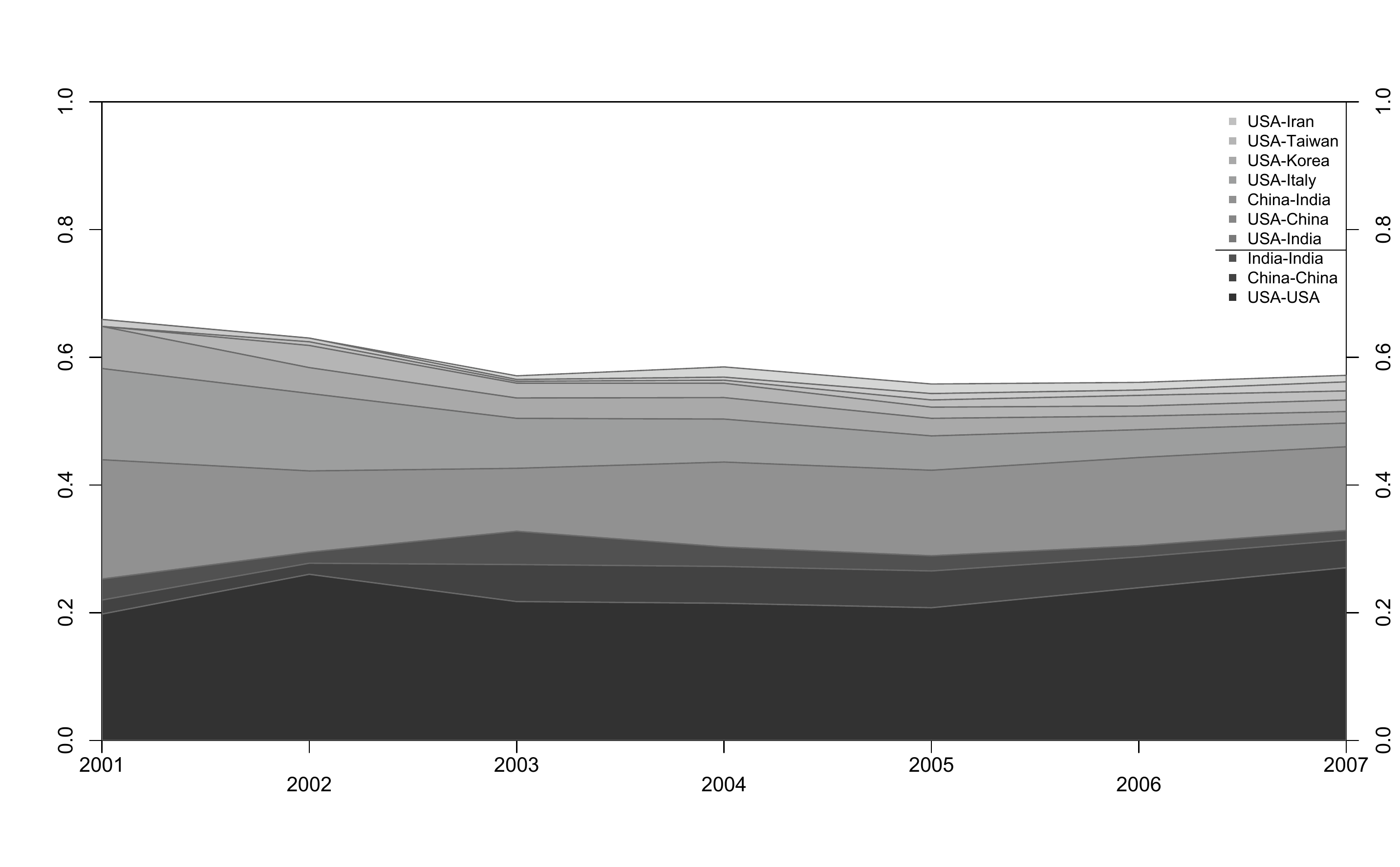}
\caption{\label{fig:orig_stack} Top ten most recurrent country of origin pairs as fraction of total volume of collaboration. Darker polygons at the bottom depict intra-national collaboration, while lighter polygons depict inter-national collaboration.}
\end{figure*}

The final discrete assortativity coefficient we analyze in this article is the country of origin of the individuals in the network. Figure \ref{fig:assort} shows that this coefficient increases steadily in the first two years and then levels off in later years at a value around $r = 0.35$. What specific intra- and inter-national collaborations may account for such trend? Even if the vast majority of publications analyzed in this study are based on research performed in the United States, it is interesting to explore the tendency of individuals to collaborate with others from their country of origin, even when they are working and living abroad, or in different countries. We present in Table \ref{tab:orig}, the ten most prominent countries that contribute to mixing patterns by country of origin. Figure \ref{fig:orig_stack} depicts these values as a stacked plot and as fraction of yearly cumulative volume of collaboration. 

\begin{table*}[h]
\centering
\begin{footnotesize}
\begin{tabular}{lllllllll}
\hline
\textbf{origin-1}&\textbf{origin-2}&2001&2002&2003&2004&2005&2006&2007\\
\hline\hline
USA&USA&36&90&150&268&332&516&686\\
India&India&4&6&40&72&92&104&110\\
China&China&6&6&36&38&38&38&38\\
\hline
USA&India&34&44&68&166&214&298&332\\
USA&China&26&42&54&84&86&94&94\\
China&India&12&14&22&42&44&46&46\\
USA&Italy&-&12&16&28&28&34&46\\
USA&Korea&-&2&2&6&18&36&36\\
USA&Taiwan&2&2&2&6&16&18&36\\
USA&Iran&-&-&4&20&24&26&26\\
\hline
\end{tabular}
\end{footnotesize}
\caption{\label{tab:orig} Most recurrent country pairs (top 10 results, cumulative values). Top 3 rows show intra-national collaboration. Bottom 7 rows show inter-national collaboration.}
\end{table*}

At the network's outset, the vast majority of collaborations is among Americans and between Americans and Indian and Chinese researchers. By year 2003, the picture only changes slightly. More intra-national collaborations appear (China-China and India-India), while inter-national collaborations between USA, India and China drop. These dynamics account for the growth of overall assortativity by country of origin recorded from 2001 to 2003, and visible in Figure \ref{fig:assort}. By year 2007, the picture becomes more variegated. Intra-national collaborations among American researchers still dominate. However, a number of novel inter-national collaborations emerge, namely between USA and Italy, South Korea, and Iran. Connecting this finding with the observation made in the previous section --- that collaboration at CENS was initially ``bootstrapped'' by faculty and later continued by graduate students and younger faculty --- it is possible to infer that as soon as CENS acquired a solid research core of collaboration, by year 2004, it began to attract and involve collaborations by international graduate students, researchers and faculty.

\section{Conclusion}
A great deal of research on scientific collaboration is performed on large-scale networks, constructed from bibliographic data harvested from domain-based and institutional document repositories. While these analyses rely on great quantities of data to study the structure, evolution and similar macroscopic features of scientific collaboration patterns, they often ignore certain contextual and microscopic factors, such as the social and academic arrangements in which collaboration takes place. This is because many available bibliographic datasets contain detailed publication metadata, but very little or no data about the authors writing those publications.

In this article, we perform a longitudinal analysis of the range, configuration and topology of a small network of scientific collaboration over a 7-year period. The network presented here is constructed from the bibliographic record of CENS, a research center involved in the development and application of sensor network technologies. Given the relatively small size of the network ($N = 291$, in its largest year), we were able to manually collect additional metadata for every individual in the network studied. We used these node characteristics to explore the assortative mixing based on academic department, affiliation, position, and country of origin.

Our findings reveal that, in the period under study, the CENS collaboration network: a) becomes more assortative in terms of academic affiliation, i.e. more intra-institutional, b) becomes less assortative in terms of academic department, i.e. more inter-disciplinary, c) is not assortative in terms of academic position, i.e. collaboration patterns are not dependent on researchers' academic positions, and d) is only slightly assortative in terms of country of origin, i.e. the extent of inter-national collaboration decreases slightly over time. 

We interpreted our findings in terms of the specific components that constitute these mixing patterns, finding that a) the increase in intra-institutional collaboration is possibly caused by CENS research consolidating around its headquarter base at UCLA, completed in 2004; b) the increase in inter-disciplinarity is largely due to the shift a CENS' research agenda, to incorporate new domains, such as civil engineering and urban planning, besides the domains traditionally associated with sensor network research, i.e., computer science and electrical engineering; c) CENS research was initially dominated by collaborations between senior researchers, but as the center matured, it became more diversified and a core research component consisting of collaborating graduate students and young faculty emerged; and d) the volume of international collaboration between USA, India, and China decreased but new smaller international efforts began as the organization matured.

This qualitative explanation of our findings revealed specific small-scale patterns that a quantitative analysis of assortativity alone would have failed to uncover. We speculate that supporting social network analyses with the proposed qualitative investigation of mixing patterns can provide a deeper understanding of the dynamics that shape (and are in turn shaped) by the changing socio-academic landscape in which scientific collaboration takes place.

\end{document}